\documentclass[12pt,a4paper]{article}
\usepackage[latin1]{inputenc}
\usepackage{amsmath}
\usepackage{amsfonts}
\usepackage{amssymb}
\author{C.~D.~Fosco$^{a}$ and G.~Torroba$^{b}$\\
{\normalsize\it $^a$Centro At\'omico de Bariloche and Instituto Balseiro}\\
{\normalsize\it Comisi\'on Nacional de Energ\'\i a At\'omica}\\
{\normalsize\it 8400 Bariloche, Argentina}\\
{\normalsize\it $^b$Department of Physics and Astronomy}\\
{\normalsize\it Rutgers University}\\
{\normalsize\it Piscataway, NJ 08855, U.S.A.}\\}
\title{Noncommutative theories
  and general coordinate transformations}
\begin{document}
\maketitle
\begin{abstract}
\noindent We study the class of noncommutative theories in $d$
dimensions whose spatial coordinates $(x_i)_{i=1}^d$ can be
obtained by performing a smooth change of variables on
$(y_i)_{i=1}^d$, the coordinates of a standard noncommutative
theory, which satisfy the relation \mbox{$[y_i\,,\,y_j] = i
\theta_{ij}$}, with a constant $\theta_{ij}$ tensor. The $x_i$
variables verify a commutation relation which is, in general,
space-dependent.  We study the main properties of this special
kind of noncommutative theory and show explicitly that, in two
dimensions, any theory with a space-dependent commutation relation
can be mapped to another where that $\theta_{ij}$ is constant.
\end{abstract}
\bigskip
\newpage
\section{Introduction}\label{sec:intro}
The usual starting point for the construction of noncommutative
quantum field theories (\cite{DN,sch,szabo}) is to assume the
existence of non-trivial commutation relations between the spatial
coordinates $x_i$, $i=1,\ldots,d$. Those relations can be summarized by an
expression with the general form:
\begin{equation}\label{eq:orig}
[x_i \,,\, x_j ]_\star \;=\; i \, \Theta_{ij}(x) \;,
\end{equation}
where $\Theta_{ij}(x)$ is an antisymmetric real matrix, which
naturally encodes the noncommutative structure of the space
considered.

The usual situation corresponds to a constant $\Theta_{ij}$, but the
space-dependent case has also found some important applications (see,
for example~\cite{cornalba,gauge,otros}). Contrary to what happens
when $\Theta_{ij}$ is a constant tensor, associativity of the $\star$-product
requires some non-trivial conditions (which have the form of
differential equations for $\Theta_{ij}$) to be true~\cite{ko1}.  Even when
those conditions are met, the construction of quantum field theories
on the resulting algebras can be rather difficult. Indeed, the
$\star$-product, as well as the derivatives and integrals, essential in
any quantum field theory application, have rather cumbersome
expressions.

Closely related to the study of space-dependent
$\star$-commutators is the consideration of general changes of
variables in a noncommutative space which is defined by the
fundamental commutation relation:
$$
[y_i\,,\,y_j]_\star  \,=\, i \theta_{ij} \,,
$$
with $\theta_{ij} \equiv {\rm constant}$. The interest in considering those
changes of variables is manifold. On the one hand, it is believed that
quantum gravity may be at the origin of noncommutativity
(\cite{rivelles,madqg}). Therefore, one would expect that general
coordinate transformations should play a role in rendering some
curved-space effects more explicit.

Besides, and this is the focus of our interest in this article, it
may be possible to use a change of variables to rewrite some
particular space-dependent commutation relations as a
constant-$\theta$ relation but in terms of new variables.  This
program was developed in~\cite{Correa:2004cm}, where some
particular examples were developed at analyzed. In those examples,
a simple study of the resulting noncommutative theory was possible
since, for example, a closed expression for the $\star$-product
can be derived in terms of the standard Moyal product.  Moreover,
derivatives and integrals can also be constructed, based on the
existence of `canonical' coordinates, namely, those that have a
constant commutator. A similar approach has been applied
in~\cite{similar} to $\kappa$-Minkowski noncommutative spacetime.

In this paper we consider the same problem in more generality,
studying a space-dependent $\Theta_{ij}$, obtained by performing a
general change of coordinates (inspired by~\cite{Jackiw:2004nm})
in a theory with a constant $\theta_{ij}$.  In
section~\ref{sec:planar}, we begin by analyzing these coordinate
transformations in a planar theory, computing the resulting
$\Theta(x)$. We also introduce an integral and derivatives, and
use those constructs to write an explicit noncommutative field
theory action, some features of which shall be interpreted as
curved-space effects. Next, in section~\ref{sec:examples}, we
apply the results of the previous section to some examples in
$d=2$. We also present a generalization of the result
of~\cite{Correa:2004cm} to higher-dimensional spaces in
section~\ref{sec:dgq2}.  In section~\ref{sec:reduct} we apply the
tools of sections~\ref{sec:planar} and~\ref{sec:examples} to show
that a general space-dependent $\Theta(x)$ may be reduced to a
constant $\theta$ by a suitable change of variables, which we
construct explicitly. Finally, in~\ref{sec:concl} we present our
conclusions.

\section{Planar theories ($d=2$)}\label{sec:planar}
To begin with, we introduce two noncommutative spatial coordinates in
\mbox{$d=2$}, \mbox{$(y_1, y_2)$}, that verify the commutation
relation:
\begin{equation}\label{eq:canon}
[y_i \,,\, y_j ]_\star \;=\; i \, \theta_{ij} \;,
\end{equation}
where $\theta_{ij}$ is a constant. Since we are working in two dimensions,
we may always write
\begin{equation}\label{eq:deftheta}
\theta_{ij}\,=\,\theta\, \varepsilon_{ij}\,
\end{equation}
where $\theta$ is a parameter with the dimensions of an area. Furthermore,
the time coordinate is assumed to commute with $y_1$ and $y_2$.

We then introduce two new coordinates, $(x_1, x_2)$, by means of a
non-singular, continuous change of variables.
Following~\cite{Jackiw:2004nm} we may write it without any loss of
generality as follows:
\begin{equation}\label{eq:trafo}
x_i \;=\; y_i \,+\, \theta_{ij} {\tilde A}_j(y) \;\;\;,\;\;\; i \,=\, 1, 2 \;.
\end{equation}
The parametrization above is well-suited for the analysis of
coordinate transformations that are continuous deformations of the
identity. Indeed, the field ${\tilde A}_j$ has the role of
determining the nontrivial content of those transformations.
Besides, the expression for the change of variables in terms of a
vector field ${\tilde A}_j$ prepares the road for the introduction
of some gauge theory concepts~\cite{Jackiw:2004nm} which find a
natural place in this context.

At this point we comment on the notation: a tilde on top of a
function has been used to denote its functional dependence in
terms of $y$ variables. This will be useful later on, when we
shall have to distinguish that from the corresponding expression
of the same object as a function of $x$ (where we shall omit the
tilde), i.e., $A_j (x) \equiv {\tilde A}_j \big(y(x)\big)$.

For the transformation (\ref{eq:trafo}) to be non-singular, a
necessary condition is that the Jacobian ${\tilde J}(y)$ be different
from zero:
\begin{equation}\label{eq:defj}
{\tilde J}(y) \;=\; \Big |\frac{\partial (x_1,x_2)}{\partial (y_1,y_2)} \Big |\;\neq \; 0.
\end{equation}
We easily see that {\em in two dimensions\/} ${\tilde J}$ may be
written more explicitly as:
\begin{equation}\label{eq:jacob2}
{\tilde J}(y)\;=\; 1 \,+\,\frac{1}{2}\,\theta \, \varepsilon_{ij} {\tilde f}_{ij}(y)
\end{equation}
with
\begin{equation}
{\tilde f}_{ij} \;=\; \partial_i {\tilde A}_j(y) -
\partial_j {\tilde A}_i(y) + \theta \{ {\tilde A}_i, {\tilde A}_j \}\;.
\end{equation}
In the previous equation, the curly bracket is defined by:
\begin{equation}\label{eq:defbr}
\{ f \,,\, g \} \;\equiv\; \varepsilon_{ij}
\frac{\partial f}{\partial y_i} \frac{\partial g}{\partial y_j} \;.
\end{equation}

We are interested in studying the effect of a general non-singular
change of variables (\ref{eq:trafo}) on the commutation relation
(\ref{eq:canon}). Namely, we want to find the form of the commutation
relation satisfied by the $x$ variables, which of course will
necessarily fall under the general form:
\begin{equation}
[ x_i \,,\, x_j ]_\star \;=\; i \; \Theta_{ij}(x) \;,
\end{equation}
where $\Theta_{ij}(x)$ is determined by (\ref{eq:trafo}). In what follows,
we will find the relation between $\Theta_{ij}(x)$ and (\ref{eq:trafo}) for
a general transformation.

A quite straightforward calculation allows one to find the commutation
relation for the `new' variables ($x_i$), although in terms of the old
ones ($y_i$):
\begin{equation}\label{eq:comm1}
[ x_i \,,\, x_j ]_\star \;=\; i \, {\tilde \Theta}_{ij}(y) \;,
\end{equation}
where
\begin{equation}\label{eq:deftTheta}
{\tilde \Theta}_{ij}(y) \;=\; \theta_{ij} \,+\, \frac{1}{2} \, (\theta_{ik} \theta_{jl} - \theta_{il}
\theta_{jk}) {\tilde F}_{kl}(y) \;,
\end{equation}
with
\begin{equation}\label{eq:defF}
{\tilde F}_{ij} (y) \;=\; \frac{\partial {\tilde A}_j}{\partial y_i} -
\frac{\partial {\tilde A}_i}{\partial y_j} - i [ {\tilde A}_i , {\tilde A}_j]_\star \;.
\end{equation}
Taking advantage of the fact that $d=2$, the previous expressions
(derived in \cite{Jackiw:2004nm} and valid for any $d$) can be further
simplified. Indeed, we may write:
\begin{equation}\label{eq:tThetad2}
{\tilde \Theta}_{ij}(y) \;=\; {\tilde \Theta}(y) \; \varepsilon_{ij}
\end{equation}
where
\begin{equation}\label{eq:tTheta}
{\tilde \Theta}(y) \;=\; \theta \, \Big(  1 \, + \, \theta \, {\tilde F}(y)\Big)
\end{equation}
with
\begin{equation}
\label{eq:Fij}
{\tilde F}(y) \;=\; \frac{1}{2} \, \varepsilon_{ij} \, {\tilde F}_{ij}(y) \;=\;
\frac{\partial {\tilde A}_2}{\partial y_1}(y) - \frac{\partial {\tilde
A}_1}{\partial y _2}(y) - i [{\tilde A}_1(y) \,,\,
{\tilde A}_2(y) ]_\star \;.
\end{equation}

So far, we have dealt with an explicit expression for
${\tilde\Theta}_{ij}(y)$. Let us see now how to write, at least formally,
the rhs in (\ref{eq:comm1}) as a function of $x$.  We may use now the
coordinate transformation that is inverse to (\ref{eq:trafo}), to
write the commutation relations for the $x_i$ coordinates as a
function of the same variables.

Using the chain rule in (\ref{eq:Fij}),
\begin{equation}
\tilde F (y)\,=\,\varepsilon_{ij} \, \frac{\partial A_i }{\partial x _k}
\frac{\partial x _k}{\partial y _j}-i[A_1 (x), A_2 (x)]_\star \,.
\end{equation}
Here, $[A_1 (x), A_2 (x)]_\star$ stands for the star product on
$x_j$-space (which we construct explicitly in~\ref{ssec:const})
induced by the change of variables (\ref{eq:trafo}) between the
functions $\tilde A _j \big(y(x)\big)$.

By introducing (\ref{eq:trafo}) in this expression each time a
derivative $\partial x / \partial y$ appears, we obtain a series expansion in powers
of $\theta$. Equivalently, the sum of that series can be found by deriving
with respect to $y_k$ in (\ref{eq:trafo}) and then applying the chain
rule, to obtain:
\begin{equation}
 \label{eq:resum1}
 \Big(\delta_{il}-\theta_{ij} \frac{\partial A_j}{\partial x _l} \Big)\,
 \frac{\partial x _l}{\partial y_k}\,=\,\delta_{ik} \,.
\end{equation}
For a well-defined change of variables, this expression can be
inverted to yield
\begin{equation}
 \label{eq:resum2}
\frac{\partial x _i}{\partial y_l}\,=\, \Delta^{-1} (x)\,
\Big(\delta_{il}+\theta_{ij} \frac {\partial A_l}{\partial x_j}
\Big) \,,
\end{equation}
where
\begin{equation}
 \label{eq:det}
\Delta (x)\,\equiv \,(1-\theta \partial_1 A_2)(1+\theta \partial_2 A_1)+\theta^2
\partial_1 A_1 \partial_2 A_2
\end{equation}
is the determinant of $\delta_{il}-\theta_{ij} \frac{\partial A_j}{\partial x_l}$.

Then, a straightforward calculation leads to
$$
\Theta (x) \,=\,\tilde \Theta \big(y(x) \big)\,=\,\theta+\theta^2
\varepsilon_{ij} \Big[\frac{1}{2}F_{ij}(x)+(\Delta^{-1}-1)\,\frac{\partial
A_j}{\partial x _i}(x)+
$$
\begin{equation}
 \label{eq:Thetax}
-\Delta^{-1}\theta \, \{A_i (x), A_j(x)\} \Big]\,,
\end{equation}
which contains all the information about the effect of the change of
variables on the commutation relation. Here,
$$
F_{ij}(x) \,\equiv\, \frac{\partial A_j}{\partial x _i} (x) - \frac{\partial
A_i}{\partial x _j}(x)-i[A_i(x)\,,\,A_j(x)]_\star \,.
$$
Expression (\ref{eq:Thetax}) is very convenient when dealing with
general, formal properties of the noncommutative theory in the new
variables. However, its application to the derivation of the
transformation between old and new variables that leads to a given
$\Theta(x)$ can be quite involved.  Indeed, the commutator $[x_1,x_2]_\star$
depends on $F_{ij}$, which in turn depends on $\star$ through
$[A_1\,,\,A_2]_\star$. This leads to a highly non-linear problem, whose
solution we shall study for some particular cases.  It is easy to get
the leading term of (\ref{eq:Thetax}) in an expansion in powers of
$\theta$:
\begin{equation}\label{eq:thetaexp}
\Theta(x) \;=\; \theta \;\; \Big[1 \,+\,\theta (\partial_1 A_2 -
\partial_2 A_1) \,+\, {\mathcal O}(\theta^2)\Big] \;,
\end{equation}
which shows that the leading order is determined by the first term
(linear in $\theta$) of the Jacobian.


In the next subsection we consider the use of a change of variables
in $d=2$ under the light of deformation quantization.

Then, in~\ref{ssec:const}, we derive some consequences and
applications of (\ref{eq:tTheta}) and (\ref{eq:Thetax}), which
summarize the effect of a coordinate transformation in $d=2$, to the
construction of noncommutative quantum field theories.

\subsection{Gauge transformations}\label{subsec:gauge}
In deformation quantization, one is interested in constructing $\star$
products only up to gauge equivalence, with gauge transformations
defined as~\cite{ko1}
\begin{equation}\label{eq:gauge1}
f(y) \, \to \, D_\theta \, f (y) \, \equiv \, \Big(1+ \sum_{m \geq
1} \theta^m D_m \Big) f(y) \,,
\end{equation}
with $D_m$ denoting differential operators. Under those
transformations, the star product transforms as follows:
\begin{equation}\label{eq:gauge2}
f \star ' g \,=\,D \big(D^{-1}f \, \star \, D^{-1}g \big) \,.
\end{equation}
We will now show that these abstract gauge transformations do indeed
have an interpretation in the context of the change of variables
(\ref{eq:trafo}).

Let us concentrate in the set of infinitesimal transformations
that leave invariant the Poisson structure (\ref{eq:canon}). They
can be written as
\begin{equation}\label{eq:gaugey}
y' _i \,=\, y_i + \theta_{ij} \tilde \xi _j (y) \,
\end{equation}
with $\tilde \xi _j$ infinitesimal and $\theta_{ij}=\theta \varepsilon_{ij}$. Expanding around $y_i$
\begin{equation}
f(y_i + \theta_{ij} \tilde \xi _j )\,=\, f(y) + \theta_{ij} \tilde
\xi _j (y) \, \partial_i f (y) + \ldots \,,
\end{equation}
which is a gauge transformation like (\ref{eq:gauge1}). We have
already calculated the effect of a transformation like
(\ref{eq:gaugey}) on $\theta_{ij}\,$; we simply use Eqs.
(\ref{eq:trafo}) and (\ref{eq:deftTheta}), identifying $\tilde A
_j \equiv \tilde \xi _j$. It is clear that
$$
\tilde \Theta_{ij} (y) \,=\,\theta_{ij}+ \mathcal{O} (\tilde \xi
\,^2) \;\; \Leftrightarrow \;\; \tilde F \,=\, 0 +\mathcal{O}
(\tilde \xi \,^2) \,.
$$
In turn, this means that $\tilde \xi$ is a pure gauge $\tilde \xi
_j = \partial_j \tilde \phi (y)$.

The set of infinitesimal gauge transformations (in the sense of
Kontsevich's formula (\ref{eq:gauge1})), corresponds then to
\begin{equation}\label{eq:gaugey2}
y' _i \,=\, y_i + \theta_{ij} \frac{\partial \tilde \phi
(y)}{\partial y _j} \,.
\end{equation}
These transformations were discussed in~\cite{Jackiw:2004nm}, from
the point of view of the mapping between fluids and noncommutative
theories; there, they are identified with the diffeomorphisms
preserving the fluid volume-element.

As a simple calculation shows \cite{Jackiw:2004nm}, the effect of
(\ref{eq:gaugey2}) on $x_j (y)$ is
\begin{equation}\label{eq:gaugex}
\delta x _j (y) \,=\, -i [x_j (y) \,,\, \tilde \phi (y)]_\star \,;
\end{equation}
it corresponds to an adjoint field in the noncommutative
representation of $U(1)_\star$. From this, it follows that
\begin{equation}\label{eq:gaugeA}
\delta \tilde A _j (y) \,=\, \partial_j \tilde \phi (y)-i [\tilde
A_j (y) \,,\, \tilde \phi (y)]_\star \,
\end{equation}
and consequently
\begin{equation}\label{eq:gaugeF}
\delta \tilde F _{ij} (y) \,=\, -i [\tilde F_{ij} (y) \,,\, \tilde
 \phi (y)]_\star \,.
\end{equation}
Therefore, the subgroup of Kontsevich's transformations preserving
$\theta_{ij}$ are indeed gauge transformations with respect to $\tilde A_j$.

From (\ref{eq:gaugeF}) and (\ref{eq:tTheta}) it follows that
$\tilde \Theta _{ij}$ is not invariant under (\ref{eq:gaugey2}),
but rather
\begin{equation}\label{eq:gaugeTheta}
\delta \tilde \Theta _{ij}(y)\,=\,-i [\tilde \Theta _{ij},\tilde
\phi (y)]_\star \,,
\end{equation}
which is another representative in the class of gauge equivalent
products of $\tilde \Theta _{ij}$. This agrees with the expected
behavior of $\Theta$:
\begin{equation}
\tilde \Theta _{ij} (y') \,=\,\tilde \Theta _{ij} (y)+\delta
\tilde \Theta _{ij}(y)\,.
\end{equation}

We are interested in reducing a noncommutative space whose coordinates
$(x_j)$ have a space-dependent $\Theta(x)$ to the $\theta$-constant case in
terms of new variables $(y_j)$, with the change of coordinates inverse
to (\ref{eq:trafo}):
\begin{equation} \label{eq:transx}
y_i\,=\,x_i - \theta_{ij} A_j (x) \,.
\end{equation}
Then the previous analysis reveals that, when this is possible,
there exists an infinite set of coordinates $y_j$ with constant
$\star$-commutator, all related through (\ref{eq:gaugey2}). In
other words, (\ref{eq:transx}) is defined up to transformations
parametrized by $\phi$:
\begin{equation}
y_i \;\to \; y'_i\,=\,x_i-\theta_{ij} \Big \{A_j (x)-\Delta^{-1}
(x)\, \Big(\delta_{kj}+\theta_{jl} \frac {\partial A_k}{\partial
x_l} \Big) \frac{\partial \phi}{\partial x _k}(x) \Big\}\,,
\end{equation}
(where we used (\ref{eq:resum2})).  For small $\theta$,
\begin{equation}
y'_i\,=\,x_i-\theta_{ij} \Big ( A_j (x)-\partial_j \phi (x) \Big)
+ \mathcal O (\theta^2) \,,
\end{equation}
which means that $A$ and its transformed by $\phi$ are in the same gauge
orbit (in this limit, for commutative Abelian gauge transformations).

\subsection{Construction of the noncommutative field
  theory}\label{ssec:const}
We will now construct a field theory in the variables $x_j$. For this,
we have to define integration and derivatives (refer to~\cite{ft} for
a general operatorial approach) and we have to find an explicit
representation for the $\star$-product in the new variables $x_j$.

The first element we consider is the integration measure $d\mu$ in the
$x$-variables: we realize that it can be simply derived from the
knowledge of the coordinate transformation and its Jacobian:
\begin{equation}\label{eq:measure}
d\mu \;\equiv\; d^2y \;=\; d^2x \; |\Delta(x)| \;,
\end{equation}
where we have assumed that the metric for the $y$ coordinates is
Euclidean and we used the relation
\begin{equation}\label{eq:yx}
y_i\,=\,x_i-\theta_{ij}A_j(x) \,.
\end{equation}
We note that (\ref{eq:measure}) is consistent with the
known formula for the measure in general coordinates. Indeed, the
metric tensor in the new coordinates, $g_{ij}(x)$, is given by:
\begin{equation}\label{eq:measure1}
g_{ij}(x) \;=\; \frac{\partial y_k}{\partial x_i} \frac{\partial y_l}{\partial x_j}
\delta_{kl} \;,
\end{equation}
and a simple calculation shows that:
\begin{equation}
g(x) \;\equiv\; \det [ g_{ij}(x) ] \;=\;  [ \Delta(x)]^2 \;.
\end{equation}
Thus, the measure (\ref{eq:measure}) is also identical to
\begin{equation}
d\mu \;=\; d^2x \,\sqrt{g(x)} \;,
\end{equation}
which is the usual expression for the volume element in general coordinates.
This shows that, indeed, there is a connection between noncommutative and
gravitational effects; refer to \cite{Correa:2004cm} for further discussion on
this issue.

Next we proceed to construct derivatives. From (\ref{eq:resum2}),
\begin{equation}\label{eq:derivy1}
D_i \,\equiv\,\frac{\partial}{\partial y_i}\,=\,\frac{\partial x _j}{\partial y
_i}\frac{\partial}{\partial x _j}\,=\,\Delta ^{-1}(x) \, \Big(\delta
_{ij}+\theta_{jl} \,\frac{\partial A_i}{\partial x _l} \Big)\,
\frac{\partial}{\partial x _j}\,.
\end{equation}
To interpret here the effect of $A_j(x)$, we rewrite this
expression as an inner derivation, with the aid of (\ref{eq:yx}):
\begin{equation}\label{eq:derivy2}
D_i\,=\,i \theta^{-1}[\varepsilon_{ij}x_j,
\;]_\star+i[A_i(x),\;]_\star \,.
\end{equation}
This equation automatically verifies Leibnitz rule for $\star$ and
$\int d \mu \, D_i f(x) \,=\,0 \: \forall f$ \cite{ft}. Besides,
we see that $A_j$ plays the role of a noncommutative connection.

Finally, we have to derive a representation for the commutation relation
\begin{equation}\label{eq:thetax2}
[x_1,x_2]_\star\;=\;i \, \Theta(x) \,.
\end{equation}
Note that, up to this point, we always started from the canonical
variables $y_i$ and arrived to the new ones $x_i$ with the change of
variables (\ref{eq:trafo}).  However, in most of the practical
problems, the situation is inverse: we start from a space-dependent
relation like (\ref{eq:thetax2}) and we would like to find a change of
variables in terms of which the commutator corresponds to the case of
constant $\theta$. What the deduction in the previous subsection shows is
that every $\Theta(x)$ of the form (\ref{eq:Thetax}) can be
reduced~\footnote{Of course, after solving the non-linear problem.
  See~\ref{sec:examples} for examples.} to a constant-$\theta$ with the
inverse change of variables of (\ref{eq:trafo}).  Consequently, if we
choose the usual Moyal product for the canonical variables:
\begin{equation}
\tilde f (y) \star \tilde g (y) \, \equiv \, {\rm exp} \Big(\frac{i}{2}\theta
\varepsilon_{jk} \frac{\partial}{\partial y_j} \frac{\partial}{\partial y' _k}
\Big) \, \tilde f (y) \, \tilde g (y') \, \Big | _{y'=y} \,,
\end{equation}
for such a $\Theta(x)$, a possible $\star$-product is
\begin{equation}\label{eq:starx}
f(x) \star g(x) \,=\, {\rm exp} \Big(\frac{i}{2}\theta\,\Gamma(x,x') \Big)\,f(x) \,g(x') \Big | _{x'=x} \,,
\end{equation}
with
\begin{eqnarray}
\Gamma &\equiv&  \Delta^{-1}(x)\Delta^{-1}(x')\, \varepsilon_{jk} \Big(\delta_{js}-\theta \varepsilon_{rs} \frac{\partial A_j}{\partial x_r} (x)
\Big) \frac{\partial}{\partial x_s} \nonumber\\
&\times & \Big(\delta_{km}-\theta \varepsilon_{nm} \frac{\partial A_k}{\partial x'_n} (x') \Big) \frac{\partial}{\partial x'_m} \,.
\end{eqnarray}

Now we have all the elements to study the effect of the coordinate transformation (\ref{eq:trafo}) on a noncommutative field theory defined on $(y_1,y_2)$. For simplicity, we consider the case of a scalar field:
\begin{equation}\label{eq:defsy}
S[{\tilde \varphi}] \;=\; \int d\tau dy_1 dy_2 \, \Big[\frac{1}{2}\big(
\partial_\tau{\tilde \varphi} \star  \partial_\tau{\tilde \varphi}  + \partial_j{\tilde \varphi} \star  \partial_j{\tilde \varphi}+ m^2
{\tilde \varphi} \star   {\tilde \varphi}\big) + V_\star  ({\tilde \varphi}) \Big] \,.
\end{equation}
Since the Moyal product between the same to functions may be written as the usual commutative product plus a total-derivate term, (\ref{eq:defsy}) can be simplified to yield:
\begin{equation}\label{eq:defsy2}
S[\tilde \varphi ] \;=\; \int d\tau dy_1 dy_2 \, \Big[\frac{1}{2}\big(
(\partial_\tau \tilde \varphi )^2   + ( \partial_j \tilde \varphi )^2 + m^2
 \tilde \varphi ^2 \,   \big) + V_\star  (\tilde \varphi) \Big] \,.
\end{equation}

Then, using Eqs. (\ref{eq:measure}), (\ref{eq:derivy1}) in
(\ref{eq:defsy2}), the action in the new variables is
\begin{equation}\label{eq:defsx}
S[\varphi] \;=\; \int d\tau dx_1 dx_2 \, \vert \Delta(x) \vert ^2 \Big[-\frac{1}{2}\, \varphi(x)\big(
\partial_\tau ^2  + D_i ^2- m^2
\big) \varphi(x) + V_\star  (\varphi) \Big] \,,
\end{equation}
with the star product in $V_\star$ computed from (\ref{eq:starx}). It is worth noting that $D_i ^2$ induces a non-diagonal quadratic term in the momentum variables and a derivative coupling:
\begin{equation}
D_i ^2 \,=\, \Delta ^{-2} (x) \big(\delta_{ij}+ \theta_{jl}\, \partial_l A_i\big)\, \Big[ \big(\delta_{in}+ \theta_{nk} \,\partial_k A_i\big)\, \partial_j \partial_n + \theta_{nk}\, \partial_j \partial_k A_i \, \partial_n\Big] \,.
\end{equation}

The propagator is directly obtained by performing the change of variables in the simple expression for $\langle \tilde \varphi (y) \tilde \varphi (y') \rangle$:
\begin{equation}
\langle \varphi(x) \varphi(x') \rangle \,=\, \frac{1}{4 \pi} \, \Big[ (t-t')^2+ \big \{(x_i-x_i')- \theta \varepsilon_{ij} \big(A_j (x) - A_j (x') \big) \big \}^2\Big] ^{-1/2} \,.
\end{equation}


\section{Examples}\label{sec:examples}

We now study some particular cases, which we define in terms of
special properties of the coordinate transformation.

\subsection{The case ${\tilde A}_2 =0$}\label{ssec:a2eq0}
When one of the components of ${\tilde A}_j$ vanishes (the second,
say) we of course have \mbox{$[{\tilde A}_1,{\tilde A}_2]_\star =0$} and
the expression for ${\tilde\Theta}$ becomes:
\begin{equation}
{\tilde\Theta}(y)\;=\; \theta \, \left[ 1 + \theta {\tilde f}(y) \right]
\end{equation}
with ${\tilde f}(y) = - \frac{\partial {\tilde A}_1}{\partial y_2}$.  Equivalently, in terms of $x_i$ we have, from (\ref{eq:Thetax})
\begin{equation}\label{eq:thetaex1}
\Theta (x) \;=\; \theta \, \frac{1}{1 + \theta \, \frac{\partial A_1}{\partial x_2}}
\end{equation}

To find $\Theta(x)$, we need the change of variables that yields $y_i$ in
terms of $x_j$; this may be written as follows:
\begin{equation}
\left\{ \begin{array}{ccl}
y_1 &=& x_1 \\
y_2 &=& x_2 + \theta {\tilde A}_1(x_1,y_2) \;.
\end{array}
\right.
\end{equation}
The second line shows that, except for some particular cases, the
explicit form of the inverse transformation for $y_2$ may not be found
exactly.  However, one can always use an expansion in powers of
$\theta$:
\begin{equation}\label{eq:ypert}
y_2 \;=\; \sum_{l=0}^\infty \, \theta_l \, y_2 ^{(l)}
\end{equation}
where the first terms are given by:
$$
y_2 ^{(0)}\,=\, x_2  \;,\;\;\;y_2 ^{(1)}\,=\, {\tilde A}_1(x_1,x_2)
\;,\;\;\; y_2 ^{(2)}\,=\, {\tilde A}_1(x_1,x_2) \partial_2 {\tilde
  A}_1(x_1,x_2)\;,\;\;\;
$$
\begin{equation}
y_2 ^{(3)}\,=\, {\tilde A}_1(x_1,x_2) \Big( \partial_2 {\tilde A}_1(x_1,x_2)
\Big)^2 \,+\,\frac{1}{2} \, {\tilde A}^2_1(x_1,x_2) \,\partial^2_2 {\tilde A}_1(x_1,x_2)
\;,\;\;\;\ldots
\end{equation}
Using the previous expansion we may also write an expansion for
$\frac{\partial A_1}{\partial x_2}$, to be used in (\ref{eq:thetaex1}) to find
$\Theta(x)$:
\begin{eqnarray}
\partial_2 A_1 &=& \partial_2 {\tilde A}_1 \;+\; \theta \, \Big[ (\partial_2 {\tilde A}_1)^2 +
{\tilde A}_1 \partial_2^2 {\tilde A}_1 \Big] \nonumber\\
&+& \theta^2 \, \Big[\frac{1}{2} {\tilde A}_1^2 \partial_2^3 {\tilde A}_1 + 3
{\tilde A}_1 \partial_2{\tilde A}_1 \,\partial_2^2 {\tilde A}_1 \,+\, (\partial_2
{\tilde A}_1)^3 \Big] \nonumber\\
&+& \theta^3 \, \Big[ {\tilde A}_1^3  \big( \frac{1}{3!} \partial_2^4 {\tilde A}_1 +
\partial_2 {\tilde A}_1 \partial_2^3 {\tilde A}_1 + (\partial_2^2 {\tilde A}_1)^2 \big)
\nonumber\\
& &\;+\;\; {\tilde A}_1^2  \big(\frac{1}{2} (\partial_2^2 {\tilde A}_1)^2 + 3  \partial_2^2
{\tilde A}_1 (\partial_2 {\tilde A}_1)^2  \big) \nonumber\\
& &\;+\;\; 4 {\tilde A}_1 (\partial_2{\tilde A}_1)^2  \,\partial_2^2 {\tilde A}_1 \,+\, (\partial_2
{\tilde A}_1)^4 \Big] \;+\; {\mathcal O}(\theta^4) \;,
\end{eqnarray}
where all the field arguments and the derivatives correspond to the
$x_1$ and $x_2$ variables. For example, ${\tilde A}_1 \equiv {\tilde
  A}_1(x_1,x_2)$.
Finally, expanding in the expression for $\Theta(x)$, we see that:
\begin{equation}
\Theta(x) \;=\; \theta \; \left[ 1 \,-\, \theta \partial_2 {\tilde A}_1 \, - \theta^2 {\tilde
    A}_1 \partial^2_2 {\tilde A}_1 \,\ldots \right] \;.
\end{equation}

The power series expansion cannot be summed exactly, except for some
particular cases.  We shall consider two of them, showing how $\Theta(x)$
may be found explicitly by solving exactly for $y$ as a function of
$x$, or by a summation of the previous series.  The explicit examples
we shall exhibit stem from an ${\tilde A}_1$ which is quadratic or
linear in $y_2$, respectively. We shall, however, come back to the
general case in the conclusions.

The quadratic case corresponds to:
\begin{equation}
{\tilde A}_1(y_1,y_2)\;=\; \alpha (y_1) (y_2)^2 \,+\, \beta (y_1) y_2 \,+\,
\gamma(y_1) \;,
\end{equation}
and produces a ${\tilde\Theta}(y)$ with the form:
\begin{equation}
{\tilde\Theta}(y) \;=\; \theta \left[ 1 \,-\, \theta \, \Big( 2 \alpha(y_1)
y_2 + \beta (y_1)\Big) \right] \;.
\end{equation}
On the other hand, we know that $y_1 = x_1$ and besides $y_2$ may be obtained from
\begin{equation}\label{eq:quadeq}
x_2 \;=\; y_2 \,-\, \theta \Big[ \alpha (y_1) (y_2)^2 + \beta (y_1) y_2
\,+\, \gamma(y_1) \Big]\;,
\end{equation}
which is a quadratic equation. Inserting its solution for $y_2$, and
$y_1=x_1$ into ${\tilde\Theta}$, we see that~\footnote{There is another solution
of the quadratic equation, which yields a $\Theta$ with the opposite sign.}:
\begin{equation} \label{eq:ex1}
\Theta(x_1,x_2) \;=\; \theta \, \sqrt{(\; 1 \,-\, \theta \, \beta (x_1)\;)^2 \;-\; 4
\theta \, \alpha (x_1) \, (\;\theta \,\gamma (x_1) - x_2\;)} \;,
\end{equation}
a result which depends on both variables, $x_1$ and $x_2$, but can
still be described in terms of the canonical variables $y_1$ and
$y_2$.  Therefore, if we start from a space-dependent $\Theta(x)$ which can
be written in the form (\ref{eq:ex1}) for an adequate choice of the
functions $\alpha \, , \, \beta \, , \, \gamma$, then the change of variables
(\ref{eq:quadeq}) will reduce the problem to the $\theta$-constant case.

Note that $\Theta(x)$ may vanish on a region (a curve, in general) of the
plane. That region is defined by the equation:
\begin{equation}
\delta (x_1,x_2) \;=\; (\; 1 \,-\, \theta \, \beta (x_1)\;)^2 \;-\; 4
\theta \, \alpha (x_1) \, (\;\theta \,\gamma (x_1) - x_2\;)\;=\;0
\end{equation}
where $\delta$ is the discriminant of the quadratic equation
(\ref{eq:quadeq}). In \cite{ft} we have analyzed the physical effects of such a behavior in the noncommutativity parameter.

We conclude our analysis of the ${\tilde A}_2=0$ example by mentioning
the linear case: ${\tilde A}_1 = y_2 \, \beta(y_1)$, which leads to:
\begin{equation}
\partial_2 A_1 \;=\; \frac{\alpha(x_1)}{1 - \theta \beta (x_1)}
\end{equation}
and
\begin{equation}
\Theta(x)\;=\; \theta \Big( 1 \;-\; \theta \, \beta (x_1) \Big)\;,
\end{equation}
in agreement with~\cite{Correa:2004cm} (after making the
identification $t(x_1) \equiv  1 \;-\; \theta \, \beta (x_1)$), and with
the proper limit of the quadratic case.
We will consider, in the next section, a generalization of this result to
$d > 2$.

It is worth noting that this kind of change of variables can be
extended to more general cases. Indeed, it is sufficient to have the
possibility of solving explicitly the equation for $x_2$ in terms of
$y_i$, and that can be done for many polynomial transformations.
Besides, we note that any polynomial transformation may always be
equivalently written as a polynomial (of the same degree) in the
algebra, when that is required.  This follows from the repeated
application of the property:
\begin{equation}
\alpha(y_1) \, (y_2)^n \;=\; \alpha(y_1) \star (y_2)^n \,-\, \sum_{l=1}^n
\, \left( \begin{array}{c} n\\ l \end{array} \right) \,
(i \frac{\theta}{2} ) ^l \, \alpha^{(l)}(y_1) \, (y_2)^{n-l}
\end{equation}
valid for all $n \in {\mathbb N}$. The resulting `$\star$-polynomial' shall
be real, since it should be equivalent to a real function (the
polynomial which only involves commutative products).

\subsection{The case $\varepsilon_{ij} \partial_i {\tilde A}_j = 0$}\label{ssec:tfeq0}
We shall assume here that ${\tilde A}_j$ verifies
\begin{equation}\label{eq:fteq0}
\varepsilon_{ij}\partial_i {\tilde A}_j (y) \;=\; 0
\end{equation}
for all the points in the plane, except for the origin $y_i = 0$. We
have in mind an Abelian vortex-like configuration for the vector field
${\tilde A}_j$; then, for the change of variables we shall assume the
domain of definition for the $y$ variables to be contained in
${\mathbb R}^2 - \{(0,0)\}$.  We can write locally the gauge field as
the gradient of a function ${\tilde \varphi}$, namely,
\begin{equation}
{\tilde A}_i \;=\; \partial_i {\tilde \varphi}(y) \;,
\end{equation}
where ${\tilde \varphi}$, to have a vortex configuration, has to be a
multivalued function.  The function ${\tilde \Theta}(y)$ will be given by
the expression:
\begin{equation}\label{eq:ex2}
{\tilde \Theta}(y) \;=\; \theta  \Big( 1 \,-\, i \, \theta
[ {\tilde A}_1 (y) \,,\, {\tilde A}_2 (y) ]_\star \Big) \;.
\end{equation}
Since the vortex is located at the origin, we fix the ${\tilde \varphi}$
function to be proportional to the polar angle:
\begin{equation}\label{eq:defvarphi}
{\tilde \varphi}(y) \;=\; \frac{q}{2\pi} \, {\rm arctan}(\frac{y_2}{y_1})
\;,
\end{equation}
where $q \in {\mathbb Z}$ is the `charge' (i.e., vorticity) of the
configuration.

Let us now consider the equations for the change of variables under
the previous assumptions
\begin{equation}
\left\{
\begin{array}{ccl}
x_1 &=& y_1 \,+\,\theta\, \displaystyle{\frac{\partial {\tilde \varphi}}{\partial y_2}}(y) \\
x_2 &=& y_2 \,-\,\theta\, \displaystyle{\frac{\partial {\tilde \varphi}}{\partial y_1}}(y)
\end{array}
\right.
\end{equation}
or, by taking (\ref{eq:defvarphi}) into account,
\begin{equation}\label{eq:trafovort}
\left\{
\begin{array}{ccl}
x_1 &=& y_1 \,-\, \displaystyle{\frac{q \theta}{2 \pi}}\,
\displaystyle{\frac{y_1}{(y_1)^2 + (y_2)^2}} \\
x_2 &=& y_2 \,-\,
\displaystyle{\frac{q \theta}{2 \pi}\,\frac{y_2}{(y_1)^2 + (y_2)^2}} \;.
\end{array}
\right.
\end{equation}
These can be more easily represented (and inverted) by introducing
polar coordinates:
\begin{eqnarray}\label{eq:defpolar}
x_1 &=& R \,\cos \phi \;\;\;\; x_2 = R \,\sin \phi \nonumber\\
y_1 &=& r \,\cos \alpha \;\;\;\; y_2 = r \,\sin \alpha  \;,
\end{eqnarray}
since (\ref{eq:trafovort}) yields:
\begin{eqnarray}\label{eq:polartrafo}
\phi &=& \alpha \;\;\;\;\;\; (0 \leq \alpha  < 2 \pi) \nonumber\\
\;\;\;
R &=& r - \displaystyle{\frac{q \theta}{2 \pi}} r^{-1} \;.
\end{eqnarray}
Regarding the range of the variables $r$ and $R$, we can distinguish
two different situations, depending on the sign of the
product $q \,\theta$. If $q \, \theta > 0$, then from
(\ref{eq:polartrafo}), we see that $R \geq 0$ requires the condition
$r \geq \sqrt{\frac{q \,\theta}{2\pi}}$ to be satisfied:
\begin{equation}
q \, \theta > 0 \;\;\Rightarrow \;\;\sqrt{\frac{q \,\theta}{2\pi}} \leq r
< \infty \;,\;\;\; 0 \leq R < \infty\;.
\end{equation}
An identical condition is obtained for $r$ when $q \,\theta \leq
0$, in order to have a one-to-one transformation
i.e., to satisfy $\frac{d R}{d r} \neq 0$, $\forall r$:
\begin{equation}
q \, \theta < 0 \;\;\Rightarrow \;\;\sqrt{\frac{q \, \theta}{2\pi}}
\leq r < \infty \;, \;\;\; \sqrt{\frac{2 q \,\theta}{\pi}}
\leq R < \infty \;.
\end{equation}

The inverse transformation becomes, in both cases:
\begin{equation}\label{eq:vortinv}
r \;=\; \Big( R \, + \, \sqrt{R^2 + \frac{2 q \theta}{\pi}} \Big)
\end{equation}
and, of course, $\alpha = \phi$.

Let us consider now the expression for ${\tilde \Theta}$ for the example
at hand. From (\ref{eq:ex2}), we see that
\begin{equation}
{\tilde \Theta}(y) \;=\; \theta  \Big( 1 \,+\, i \, \frac{q^2
\theta}{(2\pi)^2} \; [ \frac{y_2}{(y_1)^2 + (y_2)^2} \,,\,
\frac{y_1}{(y_1)^2+(y_2)^2}]_\star \Big) \;.
\end{equation}
The leading term on the rhs is determined by the Poisson bracket of the
corresponding elements. This yields, for small $\theta$:
\begin{equation}
{\tilde \Theta}(y) \;=\; \theta  \Big[ 1 \,+ \, \frac{1}{(2\pi)^2}
\; \frac{q^2 \, \theta^2 }{\big( (y_1)^2+(y_2)^2 \big)^2} \,+\,
{\mathcal O}(\theta^4) \Big] \;,
\end{equation}
or, by using (\ref{eq:vortinv}):
\begin{equation}
\Theta(x) \;=\; \theta  \Big[ 1 \,+ \, \frac{1}{2^4 (2\pi)^2} \;
\frac{q^2 \, \theta^2 }{\big( (x_1)^2+(x_2)^2 \big)^2} \,+\,
{\mathcal O}(\theta^4) \Big] \;.
\end{equation}


\section{A higher dimensional example}\label{sec:dgq2}
We will now deal with $d > 2$, showing first some of the general
features that survive from the $d=2$ case, and then considering an
example.

Our starting point is the formula for ${\tilde\Theta}_{ij}(y)$ in
$d$ dimensions:
\begin{equation}
{\tilde \Theta}_{ij}(y) \;=\; \theta_{ij} \,+\, \frac{1}{2} \, (\theta_{ik}
\theta_{jl} - \theta_{il} \theta_{jk}) {\tilde F}_{kl}(y) \;.
\end{equation}
In general, all the properties described in section~\ref{sec:planar}
are valid, except those relying on the explicit form $\theta_{ij}=\theta
\varepsilon_{ij}$. In particular, the construction of the field theory in $d$
dimensions follows the same steps as in~\ref{ssec:const}.

As an example, we consider here the natural generalization to
\mbox{$d > 2$} of the case considered at the end of~\ref{ssec:a2eq0}.
The condition we impose on the gauge field configuration is now:
\begin{eqnarray}
{\tilde A}_i(y) &=& 0 \;,\;\;\forall i = 2,\ldots,\, d \nonumber\\
{\tilde A}_1(y) &=& \sum_{j=2}^d \, y_j \, \alpha_j (y_1) \;.
\end{eqnarray}
This leads to an ${\tilde F}_{ij}$ tensor whose only non-vanishing
components are:
\begin{equation}
{\tilde F}_{k1} \;=\; - {\tilde F}_{1k} \;=\; \alpha_k (y_1)\;.
\end{equation}
Since, for the gauge field configuration defined above $y_1 = x_1$, we
see that the answer for $\Theta_{ij}(x)$, the commutator between $x_i$ and
$x_j$, is:
\begin{equation}
\Theta_{ij}(x) \;=\; \theta_{ij} \,+\, (\theta_{ik} \theta_{j1} -
\theta_{i1} \theta_{jk}) \, \alpha_k (x_1) \;,
\end{equation}
which is a function of $x_1$ only.

The measure for integration over the $x$ variables can also be
obtained explicitly, in terms of the corresponding Jacobian:
\begin{equation}\label{eq:jacobdex}
d\mu \;=\; d^d y \;=\; d^dx \, \frac{\partial
(y_1,\ldots,y_d)}{\partial (x_1, \ldots,x_d)} \;,
\end{equation}
where
\begin{equation}
\frac{\partial (y_1,\ldots,y_d)}{\partial (x_1,\ldots,x_d)} \;=\;
\frac{1}{|1\,-\, \theta^{1i} \, \alpha_i (x_1)|}\;,
\end{equation}
as a little algebra easily shows.

\section{Reduction of the general case to canonical variables}\label{sec:reduct}

In the previous sections we considered a noncommutative theory
with canonical Moyal variables $[y_1,\,y_2]_\star\,=\,i \theta$,
and we studied the effect of a general change of variables $y_i
\to x_i$ of the form (\ref{eq:trafo}). Now we have the necessary
tools to address the inverse problem, namely mapping a
noncommutative theory with general space-dependent parameter
$\Theta (x)$ to a new theory with constant $\theta$.

In the general case, the noncommutative space is a deformation of
the classical Poisson structure:
$$
\{x_1,\,x_2\}\,=\,\Theta (x)\,\,\to\,\,[x_1,\,x_2]_{\star_K}\,=\,i
\Theta (x)
$$
with Kontsevich's star product~\cite{ko1} over $C^\infty (\mathbf
R ^2)$ functions:
$$
f \star_K g \,=\,f\,g\,+\,i
\frac{\Theta}{2}\,\epsilon_{ij}\,\partial_i f \,\partial_j
g\,-\,\frac{\Theta ^2}{8}\,\epsilon_{ij} \epsilon_{kl}
\,\partial_i
\partial_k f\, \partial_j \partial_l g\,+
$$
\begin{equation}\label{eq:kont}
 \,-\,\frac{i}{12}\Theta
\partial_j \Theta \,\epsilon_{ij} \epsilon_{kl}\, \big(\partial_i \partial_k f \,\partial_l g - \partial_k f \,\partial_i \partial_l g \big) +
\ldots
\end{equation}
This satisfies the defining axioms of a star product and thus
gives a well-defined noncommutative space~\cite{cattaneo}.

We have to construct an explicit map $x_i \to y_i$ such that the
star product $\star_K$, or more generally, a gauge equivalent
product (see Eq. (\ref{eq:gauge1}) ) denoted simply by `$\star$',
gives $[y_1,\,y_2]_\star=i \theta$, with $\theta$ a constant. From
the approach described in section \ref{sec:planar}, it follows
that this is equivalent to finding a solution ${\tilde A}_j$ to
\begin{equation}\label{eq:requ1}
\Theta\big(y_i+\theta_{ij} \tilde A _j(y) \big)\,=\,\theta
\big(1+\theta \tilde F (y) \big) \,,
\end{equation}
obtained from Eqs. (\ref{eq:trafo}) and (\ref{eq:tTheta}).

Since $\tilde F$ contains the term $[\tilde A _1, \tilde
A_2]_\star$, this is in fact an infinite-order nonlinear
differential equation. Therefore, we do not expect to get a
conclusive answer from this approach. However, as we learnt
from~\ref{ssec:a2eq0}, a possible way to simplify this is to use
the ansatz $\tilde A _2 =0$. In this case, (\ref{eq:requ1})
becomes
\begin{equation}\label{eq:requ2}
\frac{\partial x_2}{\partial y_2}(y_1, y_2) \;=\,
\theta^{-1}\,\Theta(y_1, x_2) \;.
\end{equation}
This should be regarded as a nonlinear first order differential
equation for $x_2$ as a function of $y_2$, with $y_1$ playing the
role of a parameter (no derivatives with respect to $y_1$ appear
in the equation). Its solution may be found (formally) by one
quadrature:
\begin{equation}\label{eq:requ3}
y_2 \;=\; \theta \, \int \frac{dx_2}{\Theta(y_1,x_2)}
\end{equation}
where the integral is of course indefinite, and the result is not
unique unless one imposes extra (initial) conditions.

Therefore, under adequate regularity conditions, every $\Theta(x)$
may in principle be mapped to a constant $\theta$, by using Eq.
(\ref{eq:requ3}). Among the regularity conditions is of course the
non-vanishing of $\Theta$ as a function of its arguments, what is
here clearly linked to the fact that the Jacobian is everywhere
different from zero.

In this way, the geometry defined by $\Theta(x)$ and the field
theories constructed on such a space, can be traced back to the
canonical case. Of course, it may not be possible to obtain an
analytic expression for $y_2$ in the general case; however, in the
context of deformation quantization, where $\theta \to 0$,
(\ref{eq:requ2}) can always be solved by iterations.

The explicit map (\ref{eq:requ3}) connecting a general $\Theta(x)$
to the canonical constant case should not be a surprise. Indeed,
we expect noncommutative theories to emerge as certain low energy
limits of quantum gravity; besides, we know that in two
dimensions, every metric is conformally flat, with no dynamical
degrees of freedom. Only the Euler number, of topological nature,
distinguishes different gravitational backgrounds. In fact, we
have a similar situation at the level of the noncommutative
theory: compute the integral of the difference between the
commutators
\begin{equation}
 \frac{1}{i \theta^2} \; \int d^2 y \; \Big( [ x_1 \, , \, x_2 ]_\star \;-\;
[ y_1 \, , \, y_2 ]_\star \Big) \;=\; \,\int d^2y \,  {\tilde
F}(y) \;,
\end{equation}
where the commutator between $x_1$ and $x_2$ is written as a
function of the $y$ variables. We moved the constant factors in
order to have dimensionless objects on both sides.  Note that the
object on the rhs is, for well behaved changes of variables, a
topological invariant. Indeed, the `non Abelian' term vanishes for
an ${\tilde A}_j$ which decreases sufficiently fast at infinity:
\begin{equation}
\int d^2y \, [ {\tilde A}_1 \, , \, {\tilde A}_2 ]_\star \;=\; 0
\;,
\end{equation}
(i.e., when the cyclicity of the trace is valid) and we then have:
\begin{equation}
 \frac{1}{i \theta^2} \; \int d^2 y \; \Big( [ x_1 \, , \, x_2 ]_\star \;-\; [ y_1 \, , \, y_2 ]_\star
\Big) \;=\; \,\int d^2y \,  \varepsilon_{ij}\,\partial_i {\tilde
A}_j \;,
\end{equation}
which of course, can be written as a line integral at infinity,
and thus it makes the topological invariance of the object more
explicit. Note that the integral of each separate commutator is in
general divergent, but the integral of their difference can indeed
have a well-defined, finite value, at least for a class of
${\tilde A}_j\,'s$.

As a non trivial example, the case
\begin{equation}
\Theta(x)\,=\,\theta \big[\Theta_0 (x_1)+\Theta_1(x_1)x_2 \big]
\end{equation}
is particularly interesting, because (\ref{eq:requ2}) is exactly
solvable yielding
\begin{equation}
x_1\,=\,y_1\;\;\;,\;\;\;x_2\,=\,-\frac{\Theta_0(y_1)}{\Theta_1(y_1)}+C(y_1)
\,{\rm e}^{\Theta_1 (y_1) y_2} \,;
\end{equation}
here $C(y_1)$ is an arbitrary smooth function. The same can be
done in the even simpler case $\Theta_1 =0$, obtaining the same
result as in~\ref{ssec:a2eq0}.


\section{Conclusions and outlook}\label{sec:concl}

In this paper we have examined the effect of a general coordinate
transformation on a theory with constant $\star$-commutator, with the aim
of mapping (for certain cases) a space-dependent $\Theta_{ij}(x)$ theory to
another where that object is constant.

The method was first developed for $d=2$, where we explicitly
evaluated the effect of the change of variables on both the
noncommutative space-structure and on a scalar field theory action.

An interesting feature of the present approach is that, by defining
the coordinate transformation in terms of a vector field $A_j$ as
in~\cite{Jackiw:2004nm}, one can interpret (a subgroup of)
Kontsevich's equivalence relations between $\star$-products as
noncommutative gauge transformations on $A_j$.

The examples constructed in section~\ref{sec:examples} led to the
central question of whether the reduction from the space-dependent
case to the constant-$\theta$ one is always applicable or not. And
indeed, in section~\ref{sec:reduct} we showed that this is true by
constructing an explicit map. This agrees with Darboux's
theorem~\cite{abra}, which states that given a symplectic manifold
$\mathcal M$ it is always possible to find local coordinates in
the neighborhood of any point $x \in \mathcal M$ such that the
symplectic 2-form is given by
$$
\omega\,=\, dp_i \wedge dq^i \,.
$$
For the case $\mathcal M \,=\,\mathbf R ^2$, with $\Theta (x)$
positive definite, Darboux's Theorem holds {\em globally} because:
a) the symplectic form is everywhere non-singular and b) the
domain where the symplectic form is defined allows for a global
(not just local) application of the inverse Poincare Lema, as
required by the proof of Darboux's Theorem. From the point of view
of deformation quantization, this result may be regarded as a
classical limit of our map Eq. (\ref{eq:requ3}).

\section*{Acknowledgements}
C.~D.~F and G.~T.~have been supported by a Fundaci\'on Antorchas grant.
G.~T.~is supported by the Department of Physics and Astronomy of
Rutgers.

\end{document}